\documentclass[         %
aps,                    
jpa,                    
showpacs,               
nofootinbib,            
showkeys,               %
preprintnumbers,        %
floatfix]               
{revtex4}               
\tighten
\usepackage{graphicx}

\begin{document}

\title{The Role of Nuclear Physics in Understanding the Cosmos and the Origin of Elements}


\author{A.B. Balantekin}
\address{Physics Department, University of Wisconsin, Madison WI 53706 USA}

\begin{abstract}
This popular lecture, given in the conference celebrating contributions of Akito Arima to physics on the occasion of his 80th anniversary, outlines the role of nuclear physics in understanding the origin of elements. 
\end{abstract}

\maketitle

\vskip 2cm

Revolutionary advances are taking place in astrophysics motivated by the parallel developments of precision observational instruments and the state of the art nuclear physics facilities. We are now able to directly explore phenomena in the cosmos where the relevant microphysics is nuclear and particle physics. It is fair to say that this new era started with attempts to understand the origin of the solar energy. Before the advent of quantum mechanics and nuclear physics, the origin of solar energy was a mystery. Gravitational contraction to produce light would have consumed the Sun too fast, leading to the paradox of a Sun much younger than the Earth. With the introduction of relativity and quantum mechanics, some physicists started to suspect that solar energy was nuclear in origin. But the critics were concerned that the Sun was "simply not hot enough". It was Hans Bethe who eventually figured out  the specific nuclear reactions converting four protons in the Sun into an alpha particle, describing both the pp-chain \cite{Bethe:1938yy} and the CNO (Carbon-Oxygen, Nitrogen) cycle \cite{Bethe:1939bt}. It turned out that the Sun was not indeed quite hot enough for the CNO cycle, but the pp-chain (a series of nuclear reactions starting with two protons converting into a deuteron) was sufficient to power it. (The CNO cycle is more appropriate for stars much larger than the Sun). 
 
 \vskip 0.25cm
The nuclear reactions both in the pp-chain and the CNO cycle release energy in the form of either photons or neutrinos. Photons take a very long time to come out, interacting along the way with charged particles that the Sun has in abundance. On the other hand, neutrinos interact only weakly. Trillions of  solar neutrinos go through your body every second, yet they may interact only once in your lifetime. So they zip through the Sun, yet retain the memory of the nuclear fusion reactions that produced them. After Bethe's work it seemed that by detecting those neutrinos on Earth we would learn much about stellar evolution. However, their elusiveness, which helps them to get out of the Sun so quickly, also makes it very hard to detect them. In the late 1940s, Bruno Pontecorvo figured out that chlorine in a target would be converted into argon when hit by a neutrino \cite{pontecorvo}. Argon produced this way, being a noble gas, could then be measured by chemical means. 

\vskip 0.25cm
In the early 60s, John Bahcall estimated the solar neutrino flux. It was soon realized that capture into an isobaric analog state significantly enhances the neutrino-chlorine cross-section, making the experiment rather feasible. Bahcall teamed up with Ray Davis, who had been independently investigating possible use of chlorine as a neutrino target. Bahcall and Davis wrote two back-to-back papers in the Physical Review Letters describing a program the goal of which was "...to see into the interior of a star and thus verify directly the hypothesis of nuclear energy generation.." \cite{Bahcall:1964gx,Davis:1964hf}. 

\vskip 0.25cm

Davis needed an underground site. Since neutrinos interact only weakly, cosmic ray flux incident to Earth's surface overwhelms this tiny rate. One needs to go underground to filter that background. He eventually assembled his pioneering experiment at the Homestake mine in South Dakota. So one may say that this great revolution in astrophysics started with a nuclear scientist looking at the Sun from 4800 feet underground in a modest cavern. At the time there were many skeptics who said neutrino signatures of element formation in the Sun cannot be measured with the needed precision. Nevertheless Davis persisted and eventually received the Nobel Prize for his work. 

\vskip 0.25cm

Davis' pioneering experiment at Homestake was eventually joined by several other experiments measuring different parts of the solar neutrino energy spectrum. Two different radiochemical experiments observing the conversion of gallium into germanium by solar neutrinos, SAGE 
\cite{Abdurashitov:2009tn} in Russia and Gallex in Italy (later renamed Gallium Solar Neutrino Observatory, or GNO \cite{Altmann:2005ix}) were able to detect neutrinos at lower energies, where their numbers are much more abundant than at the energies the Homestake experiment can detect. Two other experiments, Kamiokande\footnote{The original goal of this experiment was to investigate if the protons in water were unstable, decaying into photons and other particles. Photons coming from solar neutrino interactions were actually the background to the signal they were searching for. Kamiokande and several other experiments built did not see any evidence of proton instability within their range of sensitivity. So the Kamiokande experimentalists did what any sensible experimentalist would do: they switched the signal and the background! This was  fortuitous: shortly after the switch was made, in February 1987, neutrinos from a supernova at the Large Magellanic Cloud made their way to Earth and were promptly detected.} in Japan (later rebuilt  on a much larger scale and called SuperKamiokande 
\cite{Abe:2010hy}), and Sudbury Neutrino Observatory (SNO \cite{Jelley:2009zz}) in Canada, could detect the high-energy tail of the solar neutrinos in real time: Kamiokande and SuperKamiokande (SK) use regular (but much purified) water as the target whereas SNO used heavy water. Neutrinos at the high energy tail of the spectrum carry about 20 times as much energy as the rest mass of electron. When these neutrinos hit the electrons in water, they make them move faster than light itself does in water. If a charged particle in a medium, such as electron, moves faster than light does in that medium, it emits light primarily in forward direction, called Cerenkov radiation. Kamiokande was the first experiment, by detecting this light, which was able to deduce not only the range of the energies of the incoming neutrino, but also show that they are indeed coming from the Sun. (Radiochemical experiments  are insensitive to the direction of the incoming neutrino and, since they have to accumulate the reaction products for some time before they can extract them, measure only an integral of the signal over time and energy). 

Neutrinos come in three different kinds, or {\it flavors}, associated with charged electrons, muons, and tauons. At the solar neutrino energies, only electron neutrinos are captured either on chlorine or gallium. 
Neutrino-electron scattering can happen for all flavors, but the scattering cross-section is about six times larger for electron neutrinos than for other flavors. Consequently, Kamiokande and SK are primarily sensitive to electron neutrinos. SNO, on the other hand, can utilize the breakup of deuterons in the heavy water to measure different components of the solar flux. Converting the neutron into a proton,  electron neutrinos could break a deuteron into two protons and an electron. Neutrinos of any flavor can also break a deuteron into its ingredients, a proton and a neutron. By detecting this released neutron, SNO could also measure the total neutrino flux from the Sun. 

In all these experiments, the electron neutrino flux measured was below the prediction of the solar models. However, the total neutrino flux measured by SNO agreed with that prediction. Conclusions of these measurements were indeed remarkable. The first conclusion was that our understanding of the Sun was pretty much correct. During their evolution in the so-called main sequence, stars indeed shine using the energy produced in nuclear fusion reactions\footnote{Indeed stars display a remarkable balance between the four forces of nature. Between two protons (or a proton and a nucleus) there is a  repulsive electromagnetic interaction and an attractive strong interaction. These two forces form a potential barrier. At stellar temperatures the system is well below the barrier maximum. Hence the reaction starts with quantum mechanical tunneling. Formation of the neutrino is via the weak interactions. Finally, the energy released in the nuclear fusion counterbalances gravity.}. 
There is still work to be done, e.g. it is essential to refine further the nuclear physics input into the solar model  \cite{Adelberger:1998qm} to assess any possible contributions of the CNO cycle to the Sun. 
 
The second conclusion of those solar neutrino experiments was that, although neutrinos are produced in the core of the Sun as electron neutrinos, they change their flavors as they travel towards the Earth. Such a transformation, also predicted by Pontecorvo, requires neutrinos of a given flavor to be a combination of neutrinos with different masses. During the last five decades, physicists showed that the so-called Standard Model of the particle physics describes all particle interactions so far measured. This model is defined by its symmetries, which happen to disallow a neutrino mass. Theoreticians developed a procedure (called effective field theory) to estimate the impact on low energy phenomena of the interactions that manifest themselves openly at higher energies yet to be reached at the accelerators (but that would have been dominant in the very early Universe). This procedure classifies the effective interactions to be added to the Standard Model by their mass dimension. Standard Model "Lagrangian" has mass dimension four. The simplest neutrino mass term one can write has mass dimension five. In fact, such a term is the only dimension-five operator that is consistent with all the symmetries of the Standard Model and its form was predicted long time ago, soon after neutrino was introduced, by Ettore Majorana \cite{Majorana:1937vz}. 

Binding energies of nuclei increase with increasing mass number up to iron. Up to iron, fusion of nuclei release energy; hence as they compensate the gravitational push, stars also produce elements up to and including iron. But the iron-group nuclei are the most tightly bound objects, hence beyond iron,  fusion requires energy intake instead of releasing it.  Consequently, beyond the iron peak another process beyond fusion is needed to form heavier nuclei. One way to produce most of the elements heavier than iron is via rapid neutron capture process (r-process). This process requires a neutron-rich site. A seed nucleus with charge $Z$ (say iron), captures neutrons in rapid succession, eventually becoming unstable against beta decay, decaying into a nucleus with  charge $Z+1$. The observed abundances of the r-process produced nuclei at distant iron-poor (i.e. first generation) stars fit the solar system abundance well for nuclei with $A>100$, suggesting a universal mechanism. To understand the r-process one needs to first understand beta-decays of nuclei both at and far-from stability: In fact, understanding the spin-isospin response\footnote{Isospin is the symmetry that transforms neutrons into protons. When the most general (relativistic) form of the weak interaction is reduced to a non-relativistic form  appropriate for heavy nuclei, one finds that in many cases the interaction also flips the spin of the nucleons.} of a broad range of nuclei to a variety of probes is crucial for not only for probing the r-process, but also for other astrophysics applications. 

Where in the Cosmos does the r-process take place? Clearly abundances of the nuclei produced in the r-process  should depend very strongly on the neutron-to-proton ratio of the environment: the larger this ratio more easily the r-process proceeds. One possible such site, suggested by Margaret Burbidge,  Geoffrey Burbidge, Willie Fowler and Fred Hoyle in their seminal paper in late 50s \cite{Burbidge:1957vc}, is the core-collapse supernovae. 

Once most of the protons in a shell are converted into alpha particles, energy production in this process stops. Gravity briefly takes over, shrinking the shell until its temperature raises sufficiently so that the helium nuclei ignite\footnote{Since there is no stable nucleus with A=8, three alpha particles need to combine to form a carbon nucleus. Because the probability of having three particles in the vicinity of each other is typically very small, in most cases this would be a very slow reaction. It is enhanced by the presence of a state in the carbon nucleus exactly in the right place for a resonance. The presence of such a state was predicted by Hoyle and the state was experimentally confirmed by Fowler.}. Once helium is exhausted in the shell this process repeats. Eventually at the core of the star, energy release by the nuclear fusion processes stop. At that point there is nothing to prevent gravity from asserting itself. 
The core collapses until the Pauli exclusion principle, which prevents two fermions from occupying the same state, stops the collapse. An outgoing shock wave is formed. Once this shock wave hits the envelope of the star, it ignites iron-group (primarily cobalt and nickel) nuclei there, producing the supernova light, which would briefly outshine the galaxy. However mighty  those fireworks may have seemed to our ancestors, they are still a side show to the main event. This light only carries about one percent of the total energy released in the explosion. 

This collapse is a very orderly (low-entropy) event. For smaller stars (with masses 8 to 12 $M_{\odot}$)\footnote{ $M_{\odot}$ stands for the solar mass.}
this collapse already occurs from the oxygen-neon-magnesium core, as the star can never get hot enough to ignite this core. For heavier stars (with masses more than 12 $M_{\odot}$) iron can be formed in the core before it collapses. During the collapse, electrons are captured onto protons in the nuclei, forming neutrons. By the time the collapse stops, one has a compact object of mostly neutrons extending a mere ten kilometers or so. This "proto-neutron star" is very hot: it carries 99\% of the gravitational binding energy of the pre-supernova star (about $10^{53}$ ergs). The easiest way for it to shed energy is to release this energy in the form of neutrino-antineutrino pairs. Altogether $10^{58}$ neutrinos are emitted. The main event in a core-collapse supernova is converting essentially the entire gravitational binding energy of the massive pre-supernova star into neutrinos!  

\vskip 0.25cm
Burbidge {\it et al.} suggested the neutron-rich ejecta outside the core in a type II supernova as a possible site of the r-process nucleosynthesis. Clearly, as this site is bombarded by the neutrinos emitted by the cooling neutron star, neutrinos play a special role in the dynamics of core-collapse supernovae and the r-process nucleosynthesis they may host \cite{Balantekin:2003ip}. Neutrinos streaming out of the neutron star control the neutron-to-proton ratio, which is the controlling parameter of the r-process nucleosynthesis. For example, a black hole formation at the center of the star would truncate the neutrino emission. Evidence of such a truncation may exist in the observed abundances of the nuclei produced in the r-process \cite{Sasaqui:2005rh}. More importantly, the neutrino gas in the vicinity of the neutron star, unlike any other situations with neutrinos, will be dominated by the neutrino-neutrino interactions. Much attention has been paid to the role of neutrino-neutrino interactions in the evolution of this neutrino gas (for recent reviews see references \cite{Duan:2009cd}, \cite{Duan:2010bg}, and 
\cite{Raffelt:2010zza}). In particular, the impact of neutrino-neutrino interactions on the r-process nucleosynthesis yields are still being explored \cite{Balantekin:2004ug}. 

\vskip 0.25cm

Understanding element formation necessitates a careful study of nuclear structure, nuclear dynamics and underlying symmetries. In this conference we are celebrating the contributions of Akito Arima to the development of nuclear physics as well as to science policy on the occasion of his 80th birthday. Arima has contributed to many aspects of nuclear physics crucial to our understanding of the Cosmos. For example he developed the Interacting Boson Model in collaboration with Franco Iachello 
\cite{Arima:1975zz}. Many of his contributions are described by other contributions to these proceedings. 

One should not be left with the impression that we understand everything about the origin of the elements. As we delve deeper into the mysteries of the Universe we always come up with new puzzles. 
One example is the wealth of data we recently gathered about the cosmic microwave background radiation. Observation of these photons from the Early Universe provides us an estimate of the elements produced in the Big Bang. One such element made in the Early Universe is an isotope of lithium, $^7$Li. 
Low-metallicity\footnote{Astronomers typically call all the elements heavier than helium metals. Low metallicity in a star implies that the material in it has been previously processed, indicating a star formed not much after the Big Bang.} halo stars exhibit a plateau of  $^7$Li abundance, indicating their  primordial origin \cite{Spite:1982dd}. However, the amount of $^7$Li needed to be consistent with the microwave photon observations is significantly more than $^7$Li observed in old halo stars. The origin of this inconsistency is an open question. Clearly much more work needs to be done to fully understand the origin of the elements. 

Before concluding these remarks, I would like to touch upon another of Arima's intellectual pursuits; he is a well-known haiku poet. Many of us who attended physics conferences with him remember the little black book he carries to jot down his poems. Often one finds Akito Arima the physicist in the writing of Akito Arima the poet. For example in his book {\it Einstein's Century} \cite{arimahaiku}, he extols the end of the twentieth century that started with many of Einstein's accomplishments: 
\begin{verse}
The dog star: \\
Einstein's century \\
comes to an end
\end{verse}
One other poem in this book is about spring in the Golden Hall Temple in Hiraizumi in the northeastern Tohoku region of Japan. This poem was inspired by an earlier poem by the haiku master Matsuo Basho, also about the same temple. Basho wrote many haikus about this region, his contemplation that only the grass remains in the fields where the warriors of a bygone era once fought is well-quoted. 
\begin{figure}
\includegraphics[height=.4\textheight]{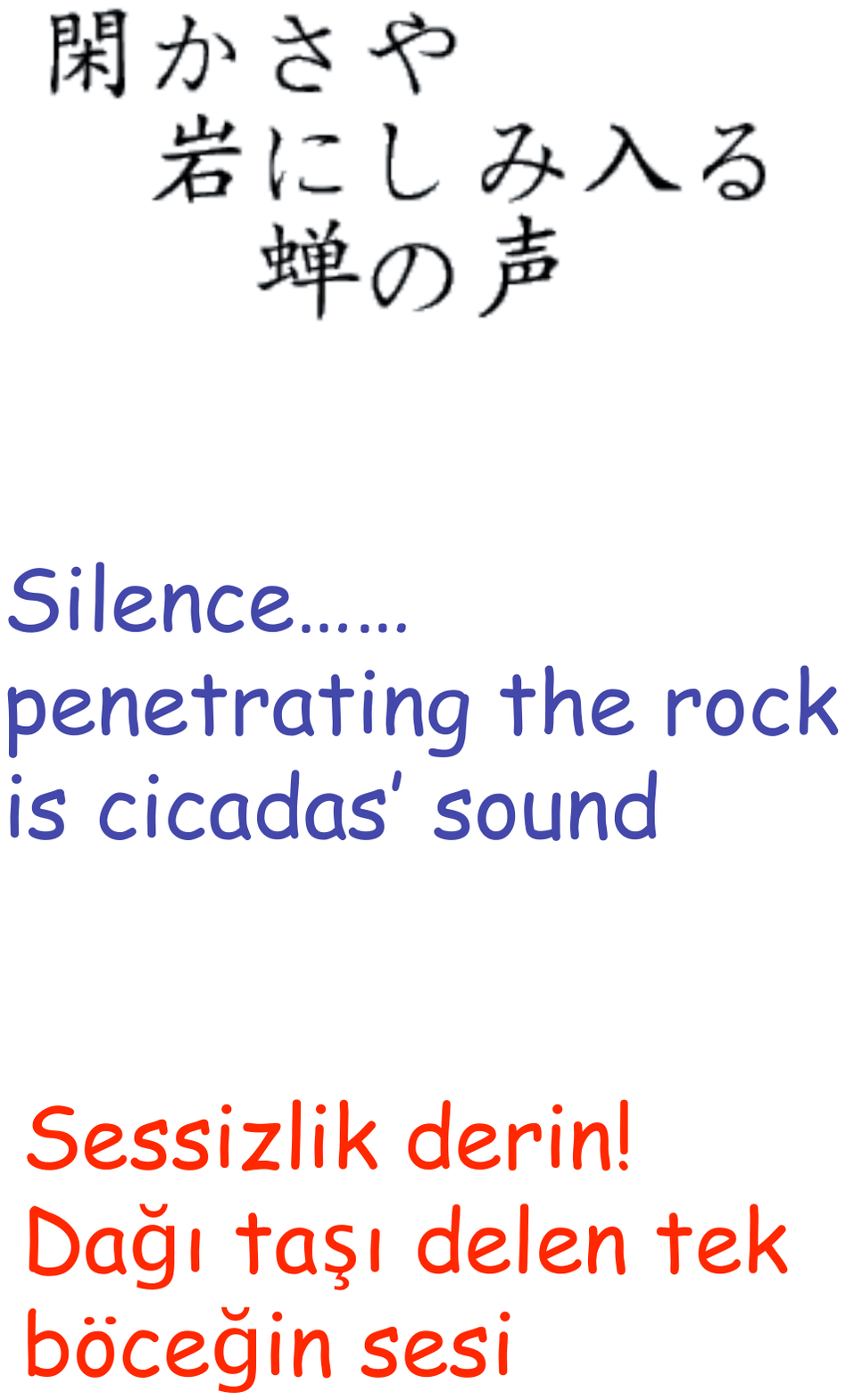}
 \caption{Basho's poem about Yamadera in the original Japanese with English and Turkish translations. Observers of symmetries would note that the Turkish translation keeps the haiku meter of 5-7-5 syllables.}
\end{figure} 
However another, equally well-known haiku by Basho is probably more suited to our subject, the one he wrote at another temple, Yamadera (see Figure 1). In this poem Basho notes that silence is broken by the sound of cicadas in the fields. It is often said that a wave of comforting nostalgia sweeps over  people who grew up with that sound when they hear cicadas. To them listening to {\it semi no koe}\footnote{The sound of Cicada} is like listening to a melody of Nature. Physicists feel a similar joy listening to what neutrinos tell us about the Universe!


\section*{Acknowledgments}
This work was supported in part
by the U.S. National Science Foundation Grant No. PHY-0855082
and
in part by the University of Wisconsin Research Committee with funds
granted by the Wisconsin Alumni Research Foundation.



\begin{thebibliography}{99}

\bibitem{Bethe:1938yy}
  H.~A.~Bethe and C.~L.~Critchfield,
  \emph{Phys.\ Rev.}  \textbf{54}, 248--254 (1938).
  
\bibitem{Bethe:1939bt}
  H.~A.~Bethe,
  \emph{Phys.\ Rev.}  \textbf{55}, 434--456 (1939).

\bibitem{pontecorvo}
B. Pontecorvo, "Inverse Beta Decay" \emph{National Research Council of Canada, Division 
of Atomic Energy,Chalk River, Report PD-205} (1946) (reproduced in 
\url{http://pontecorvo.jinr.ru/work/ibp.html}). 

\bibitem{Bahcall:1964gx}
  J.~N.~Bahcall,
  \emph{Phys.\ Rev.\ Lett.}  \textbf{12}, 300--302 (1964).

\bibitem{Davis:1964hf}
  R.~Davis,
  \emph{Phys.\ Rev.\ Lett.}  \textbf{12}, 303--305 (1964).
  
\bibitem{Abdurashitov:2009tn}
  J.~N.~Abdurashitov {\it et al.}  [SAGE Collaboration],
  \emph{Phys.\ Rev.\  C} \textbf{80}, 015807 (2009)
  [arXiv:0901.2200 [nucl-ex]] and references therein.

\bibitem{Altmann:2005ix}
  M.~Altmann {\it et al.}  [GNO COLLABORATION Collaboration],
  \emph{Phys.\ Lett.\  B} \textbf{616}, 174--190 (2005)
  [arXiv:hep-ex/0504037] and references therein.

\bibitem{Abe:2010hy}
  K.~Abe {\it et al.}  [Super-Kamiokande Collaboration],
  arXiv:1010.0118 [hep-ex].
  
\bibitem{Jelley:2009zz}
  N.~Jelley, A.~B.~McDonald and R.~G.~H.~Robertson,
  \emph{Ann.\ Rev.\ Nucl.\ Part.\ Sci.}  \textbf{59}, 431--465 (2009) and references therein.
  
\bibitem{Adelberger:1998qm}
  E.~G.~Adelberger {\it et al.},
  \emph{Rev.\ Mod.\ Phys.}  \textbf{70}, 1265--1292 (1998)
  [arXiv:astro-ph/9805121]; 
\bibitem{Adelberger:2010qa}
  E.~G.~Adelberger {\it et al.},
  arXiv:1004.2318 [nucl-ex].
  
\bibitem{Majorana:1937vz}
  E.~Majorana,
  \emph{Nuovo Cim.}  \textbf{14}, 171--184 (1937).
  
\bibitem{Burbidge:1957vc}
  M.~E.~Burbidge, G.~R.~Burbidge, W.~A.~Fowler and F.~Hoyle,
  \emph{Rev.\ Mod.\ Phys.}  \textbf{29}, 547-650 (1957).
  
\bibitem{Balantekin:2003ip}
  A.~B.~Balantekin and G.~M.~Fuller,
  \emph{J.\ Phys.\ G} \textbf{29}, 2513--2522 (2003)
  [arXiv:astro-ph/0309519].

\bibitem{Sasaqui:2005rh}
  T.~Sasaqui, T.~Kajino and A.~B.~Balantekin,
  \emph{Astrophys.\ J.}  \textbf{634}, 534-- 541(2005)
  [arXiv:astro-ph/0506100].
  
\bibitem{Duan:2009cd}
  H.~Duan and J.~P.~Kneller,
  \emph{J.\ Phys.\ G} \textbf{36}, 113201 (2009)
  [arXiv:0904.0974 [astro-ph.HE]].

\bibitem{Duan:2010bg}
  H.~Duan, G.~M.~Fuller and Y.~Z.~Qian,
  \emph{Annu. Rev. Nucl. Part. Sci}. \textbf{60}, 569 (2010) [arXiv:1001.2799 [hep-ph]].

\bibitem{Raffelt:2010zza}
  G.~G.~Raffelt,
  \emph{Prog.\ Part.\ Nucl.\ Phys.}  \textbf{64}, 393--399 (2010).

\bibitem{Balantekin:2004ug}
  A.~B.~Balantekin and H.~Yuksel,
  \emph{New J.\ Phys.}  \textbf{7}, 51 (2005)
  [arXiv:astro-ph/0411159].

\bibitem{Arima:1975zz}
  A.~Arima and F.~Iachello,
  \emph{Phys.\ Rev.\ Lett.} \textbf{35}, 1069--1072 (1975).
 
\bibitem{Spite:1982dd}
  F.~Spite and M.~Spite,
  \emph{Astron.\ Astrophys.}  \textbf{115}, 357--366 (1982).

\bibitem{arimahaiku}
A. ~Arima, \emph{Einstein's Century: Akito Arima's Haiku}, Brooks Books, Decator, IL, 2001 
(ISBN: 1-929820-01-1) . 


\end{thebibliography}
\end{document}